# Run II Searches for Supersymmetry

David Toback

*Texas A&M University*
*For the CDF and DØ Collaborations*

**Abstract.** In these proceedings, we present all of the Run II results in the search for Supersymmetry from the Tevatron. At the time of presentation ~200 pb$^{-1}$ of Run II data had been analyzed for the presence of new physics. We summarize searches for CHAMPS, Squarks and Gluinos in the jets+Missing Transverse Energy ($\not{E}_T$) final state, multiple dilepton final state analyses including $B_S \to \mu\mu$, RPV Sneutrinos and Chargino/Neutralino production, and conclude with GMSB searches in the $\gamma\gamma + \not{E}_T$ final state. While there is no evidence of new physics, we do note some interesting events and modest excesses for future reference as more data comes in. In many cases, the limits presented are the worlds first or most stringent.

## INTRODUCTION

In these proceedings we summarize all of the Run II results from the first 200 pb$^{-1}$ of data analyzed in the search for Supersymmetry from the Tevatron. Most models focus on mSUGRA/MSSM, GMSB and/or RPV scenarios. These include new results on searches for CHAMPS, searches for Squarks and Gluinos in the jets+Missing Transverse Energy ($\not{E}_T$) final state, searches in the dilepton final state, and GMSB searches in the $\gamma\gamma + \not{E}_T$ final state. While there is no evidence of new physics, we do provide some details on some interesting events and modest excesses where attention should be paid as more data comes in. In many cases, these results are the new world's best limit and/or "hot off the presses" result (at least at the time of the presentation).

## CHARGED MASSIVE STABLE PARTICLES (CHAMPS)

CDF has used its new Time-of-Flight (TOF) system to search for CHArged Massive Stable Particles, or CHAMPs. These particles would behave like slow, but high $P_T$ muons in the detector. A search in the first 53 pb$^{-1}$ of data investigates isolated tracks in the detector with $P_T > 18$ GeV [1]. Figure 1 shows the distribution of the time of arrival of data, in nsec, compared to backgrounds. A total of 2.9±0.3(stat)±3.1(sys) events are expected from instrumental mis-measurement with a time-of-flight above 2.5 nsec. A total of 7 events are observed in the data. While any excess is noteworthy, this one is small and not statistically significant. We further note that almost a factor of 10 more data has already been taken so this channel should be pursued as it has the potential to be quickly confirmed or excluded with greater precision.

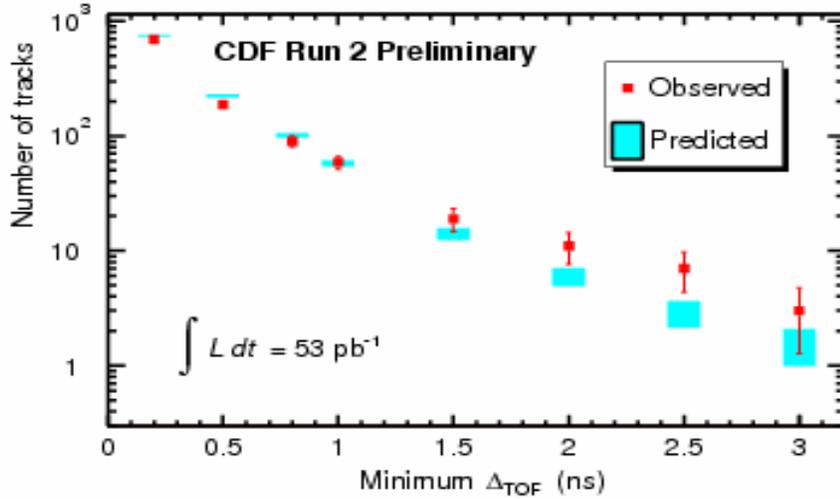

**FIGURE 1.** The time-of-flight for high $P_T$ tracks in the search for CHAMPs at CDF. The final cut is at 2.5 nsec. While there is a small excess above expectation, it is not statistically significant. There is no significant excess above background expectations. More data, already taken but not analyzed, will hopefully be able to answer the question of whether this excess is real.

Since there is no evidence for new physics, limits are set. Many theories predict CHAMPs. Within SUSY Stops, Staus, Charginos, and Sleptons could all be long lived. For theoretical reasons of production simplicity, new limits are set on Stop production. As shown in Figure 2, Stops that are below 107 GeV are excluded at 95% C.L. This is the new world limit and extends the previous best limit from ALEPH of 95 GeV. We note that this result is a nice complement to the Run I decay based searches for prompt Stop decays.

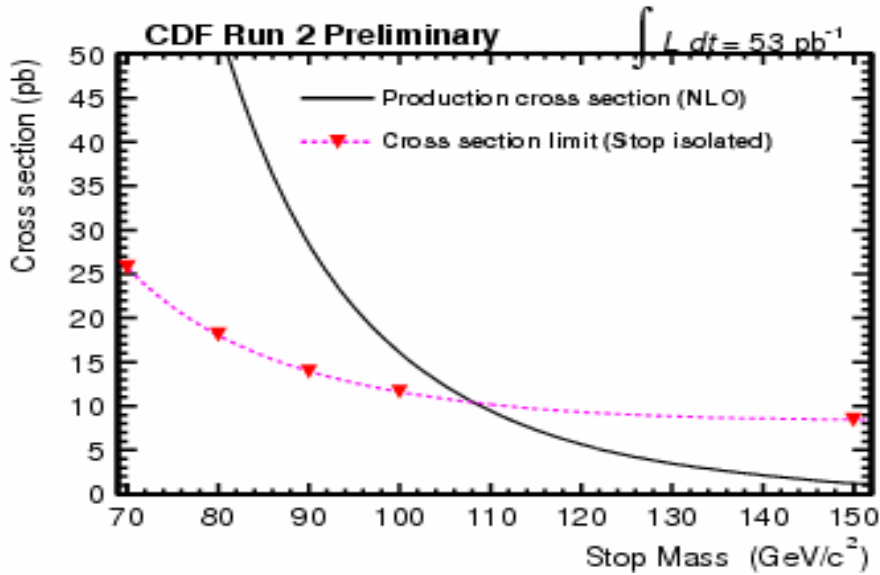

**FIGURE 2.** The 95% C.L. Stop production cross-section upper limits and a comparison to theoretical predictions at NLO. A Stop is ruled out for masses below 107 GeV at 95% C.L.

# SEARCH FOR NEW PHYSICS IN JETS+$\not{E}_T$

The Tevatron has long done direct searches for Squarks and Gluinos in the jets+$\not{E}_T$ final state. Recently, both CDF and DØ have separately produced new preliminary results in both the light-quark jets+$\not{E}_T$ mode, and the heavy flavor jets+$\not{E}_T$ mode. In both cases, the sensitivity has been extended beyond LEPs kinematic reach and provided new search sensitivity.

The light-quark jets+$\not{E}_T$ analysis searches for Gluinos and light Squarks by looking for acoplanar jets+$\not{E}_T$. This analysis is done in 85 pb$^{-1}$ of data at DØ [2]. To summarize, the analysis requires at least two large jets and Total $H_T$>275 GeV. One of the interesting features of this search is that the backgrounds are dominated by electroweak production of Z→νν+2 jets and W→τν + jets as opposed to QCD production with fake $\not{E}_T$. The data is shown in Figure 3LHS. The final cut is $\not{E}_T$>175 GeV. Above this threshold, a total of 2.67±0.95 events are expected with 4 events observed in the data. We note that in Figure 3LHS there is a single event out on the tail worthy of note, shown in Figure 3RHS. It has two big jets: $E_T$=289 GeV and 117 GeV, 2 little jets: $E_T$=14 GeV and 11 GeV and $\not{E}_T$=381 GeV. While it is an *a posteriori* estimate, we note that about 1 event is expected above 300 GeV; ~¼ above 350 GeV. It is not clear what this event is, and we note that from the plot it does not particularly look like signal or background.

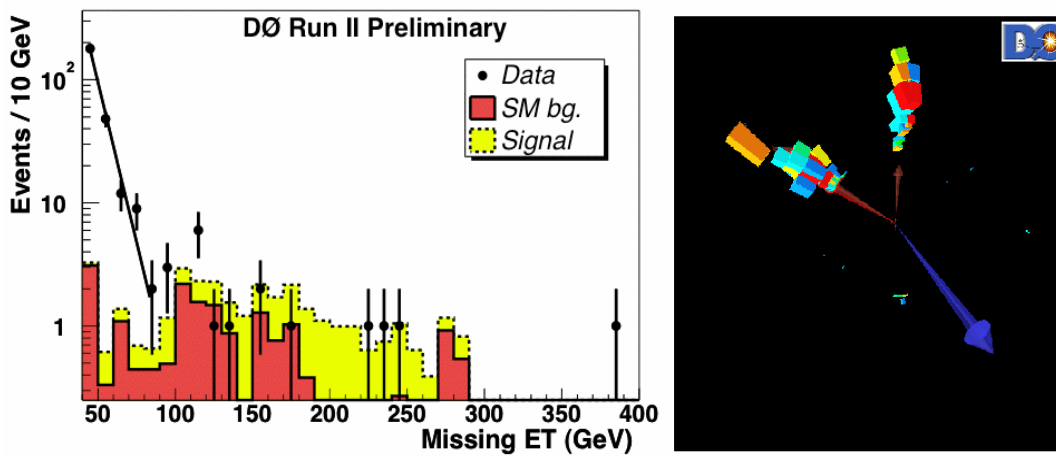

**FIGURE 3.** The LHS shows the $\not{E}_T$ distribution in the DØ acoplanar jets+$\not{E}_T$ analysis to search for Squarks and Gluinos. The dark shaded histogram is background from electroweak sources. QCD is estimate from the exponential for low values of $\not{E}_T$ as shown in the figure. The final cut is at $\not{E}_T$>175 GeV. The RHS shows an event display for the interesting event at high $\not{E}_T$ and described in the text.

Since there is no evidence for new physics we set limits on light Squarks and Gluinos in an in mSUGRA scenario ($m_0$=25 GeV, $A_0$=0, *tanβ*=3, μ<0 and varying $m_{1/2}$). The results are shown in Figure 4 with a new Gluino mass limit of 333 GeV (for Squark masses as low as they can be in mSUGRA). This extends previous world's best limit (previously set by CDF in Run I).

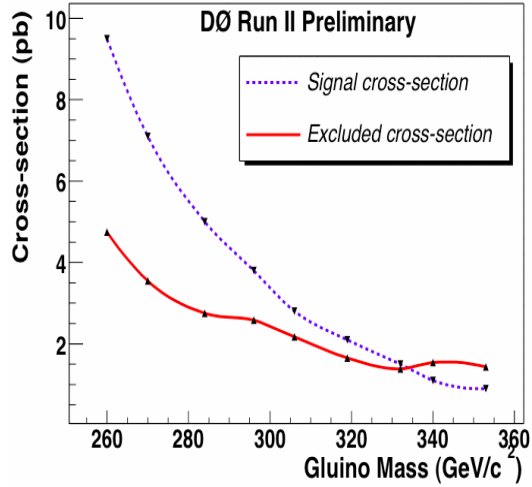

**FIGURE 4.** The 95% C.L. cross section upper limits on the light Squarks and Gluinos. The dashed line is the expected production cross section.

A second jets+$\not{E}_T$ search, this time from CDF in 156 pb$^{-1}$, requires four jets and $\not{E}_T$ but requires a b-tag [3]. This analysis is optimized for pair production and decays of Gluinos via Gluino→Sbottom+bottom. There are two separate analyses: a single-tag and a double-tag analysis. In both, the backgrounds are dominated by top quark pair production and decay. In the single-tag analysis, only one of the jets is required to be tagged yielding an expectation of 16.4±3.6 events in the data; 21 are observed. In the double-tag analysis, expected to be more sensitive, see Figure 5, a total of 2.6±0.7 events are expected with $\not{E}_T$ >80 GeV with 4 events observed in the data. From the figure we see that there is a single bin with a slight excess and note that *a posteriori* there is ~1/2 event expected above $\not{E}_T$ > 125 GeV, and 2 are observed.

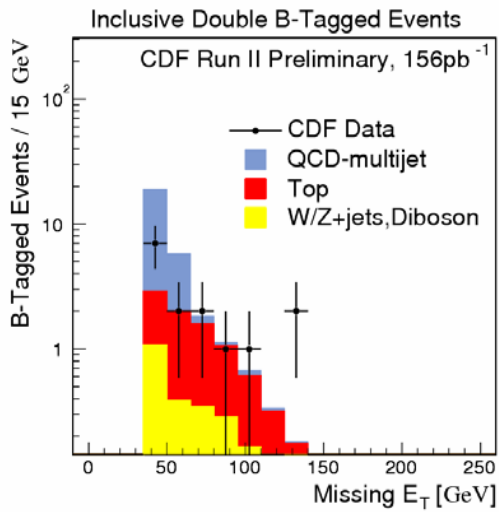

**FIGURE 5.** The $\not{E}_T$ distribution in the CDF double b-tagged jets sample in the search for Sbottom Squarks.

Since there is no evidence of new physics limits are set as shown in Figure 6. While these new limits are not sensitive to when Sbottom mass is close to the LSP mass (the LEP sensitivity is still best there) and low mass Gluinos, they significantly extend the best limits at high mass Sbottom and low mass Neutralinos.

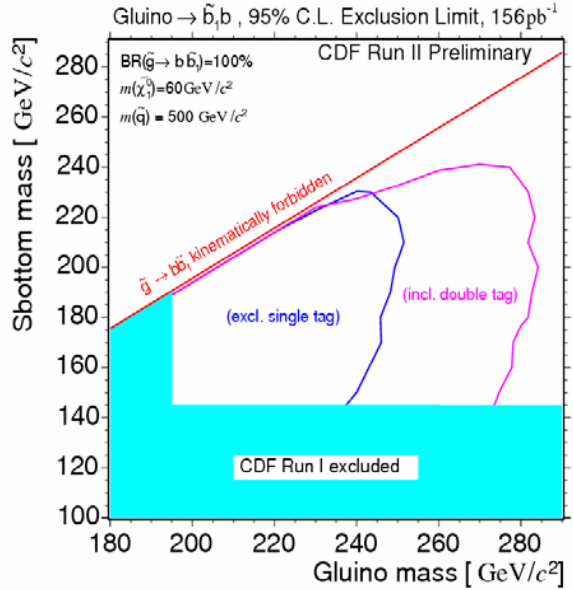

**FIGURE 6.** The 95% C.L. exclusion region from the CDF four jet+$\not{E}_T$ search for Gluino pair production and decay into Sbottom quarks. These results, for large Sbottom-Neutralino mass differences, significantly improve on previous limits.

## MULTILEPTON SEARCHES

Searches for SUSY in multi-lepton final states have been a staple of the Tevatron program for many years at both CDF and DØ. We next present three separate types of searches: Low mass indirect searches in $B_s \to \mu\mu$, a high mass resonance search for RPV Sneutrinos, and a search for Chargino/Neutralino pair production and decay.

### $B_S \to \mu\mu$

CDF has recently published an indirect search for SUSY in the $B_s \to \mu\mu$ channel in 171 pb$^{-1}$ of data [4]. Loop diagrams containing SUSY particles could affect the branching ratio by one to three orders of magnitude, potentially making this the most sensitive, albeit indirect, test of SUSY at the Tevatron. After a number of optimized topology cuts, the dimuon invariant mass is shown in Figure 7. The final selection requirement is a 4σ mass window around known world average of the $B_S$ mass. A total of 1.1±0.3 events are expected from known SM backgrounds and there is 1 event observed in the data. Using the result a new 95% C.L. branching ratio limit is set at Br($B_s \to \mu\mu$) = 7.5x10$^{-7}$. This is the new world's best limit and significantly improves on the previous limit of $2.0 \times 10^{-6}$ (CDF Run 1). We note that the SM branching ratio is predicted to be $3.5 \times 10^{-9}$.

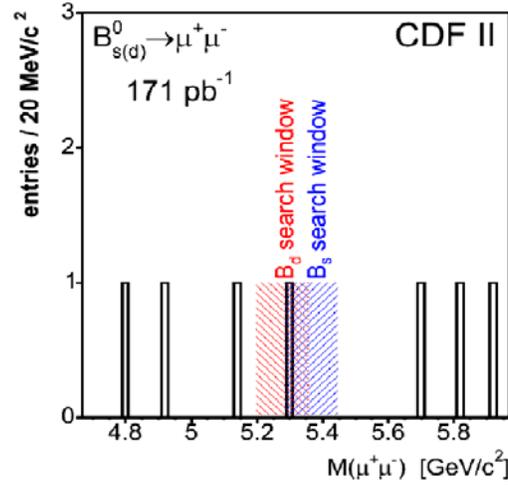

**FIGURE 7.** The μμ invariant mass in the CDF search for Bs→μμ. Note that this search the search window is 4σ around the measured $B_S$ mass. This plot is updated since the conference.

## High Mass Resonance Searches

Tevatron searches for resonances in the *ee* and μμ channels have been done for many years for Z', E6, Higgs, Technicolor etc. New results in 200 pb$^{-1}$ of data were presented in talks by M. Gold/M. Unel Karagoz, since there is no excess above SM background expectations, limits are set in a new interpretation in terms of RPV Sneutrino production and decay [5]. These are shown in Figure 8 and are the first limits for large Sneutrino masses.

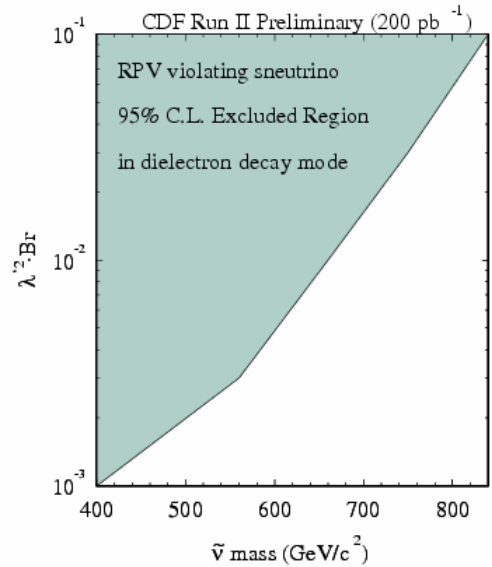

**FIGURE 8.** The exclusion region for the RPV search for Sneutrinos from the CDF *ee* and μμ final state search. These are the first limits on large mass Sneutrino production on the RPV scenario scenarios described in [5].

# Search for Chargino/Neutralino Pair Production

A gold plated signature of SUSY Chargino/Neutralino pair production and decay is the three lepton+$\not{E}_T$ final state. Since the end of Run I, LEP has significantly improved the limits. Recently DØ has produced three new results in this mode: same-sign (i.e., same electromagnetic charge) muons, two electrons + lepton, and electron + Muon + lepton. Each has a slightly different strategy and is combined in the end to produce a best limit [6].

Same-Sign Muons: The strategy here is to increase the acceptance by only requiring two out of three leptons, but reduce background by requiring them to be same-sign. The event requirements are $P_{T1}>11$ GeV, $P_{T2}>5$ GeV, $\not{E}_T>15$ GeV and $M_{\mu\mu}<80$ GeV. In 147 pb$^{-1}$ of data 0.13 ± 0.06 events of background are predicted (dominated by WZ (0.07) and bb (0.04)). One candidate is observed in the data. Since there is no third lepton candidate it is not clear what this background is, but is most likely bb.

Electron+Muon+Lepton: The analysis strategy here is to require two leptons of any charge combination, but get background rejection by the additional requirement of a third isolated track which might indicate a third lepton (e, μ or τ). Events are required to have an electron with $E_T>12$ GeV, a muon with $P_T>8$ GeV, and $\not{E}_T>15$. Figure 9 (LHS) shows the $P_T$ distribution of the third track. After a cut of 3 GeV, in 158 pb$^{-1}$ of data, a total of 0.5±0.2 is expected with 0 events observed.

Two electron+lepton: Finally, a search is done in a sample requiring two-electrons and a third track (which could be a lepton). Figure 9 (RHS) shows the distribution of the $\not{E}_T$ x $P_T$ of the third track which has been shown to be an effective separation between signal and background in this channel. After a cut >250 GeV$^2$, in 175 pb$^{-1}$ of data a total of $0.3^{+0.4}_{-0.3}$ events are expected with 1 event observed in the data. This interesting event has electron $E_T$'s of 33 GeV and 26 GeV, a track $P_T = 8.6$ GeV and $\not{E}_T = 52.1$ GeV. While the backgrounds are dominated by WW, further inspection of the event indicates it is likely a Wγ event where the photon converted.

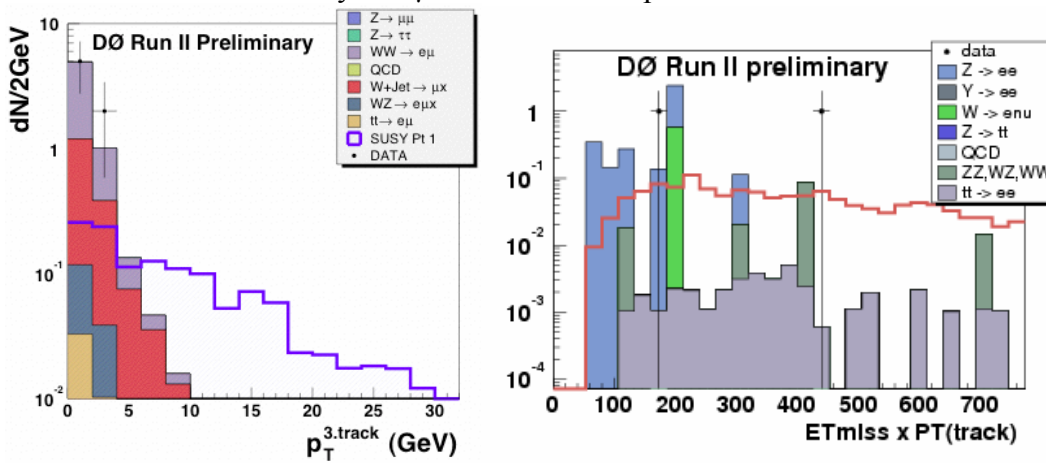

**FIGURE 9.** The LHS shows the $P_T$ spectrum of tracks in the DØ electron+muon+lepton analysis. The final selection criteria is for a third track with $P_T>3$ GeV. No candidates pass this requirement. The RHS shows the $\not{E}_T$ x $P_T$ distribution for the two electron+lepton sample from DØ. The final section requirement is for this quantity to be > 250 GeV$^2$. The one interesting event is described in the text.

Combining the results: All three searches are optimized for the region above the LEP limits. The combined results produce the most stringent limits from the Tevatron to date. Unfortunately, as shown in Figure 10, while they are a significant improvement over Run I results, there is no extension yet of the exclusion region from LEP.

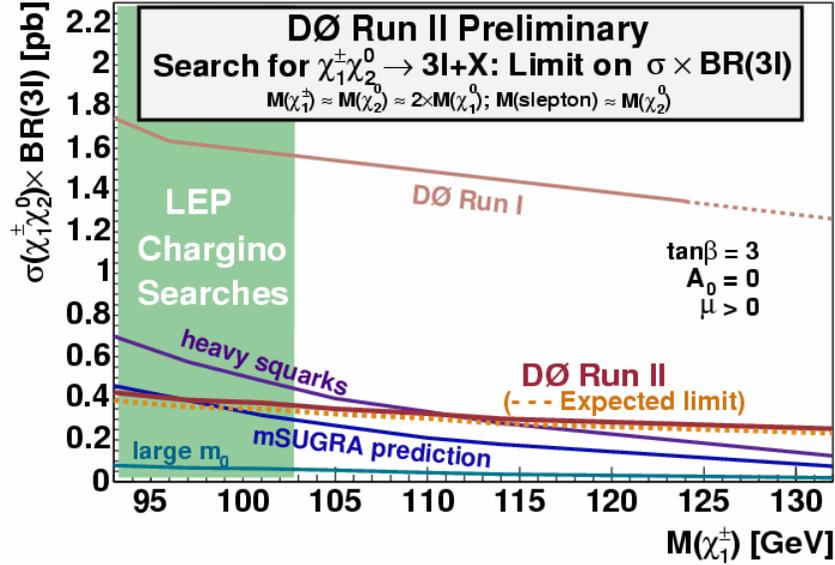

**FIGURE 10.** The combined DØ limits on the search for Chargino/Neutralino pair production and decay (updated since the conference).

## GMSB SEARCHES IN TWO PHOTONS+MET

Both the CDF and DØ collaborations have searched for new physics in the $\gamma\gamma + \slashed{E}_T$ channel [7]. This search is done both to search for other events like the CDF Run I $ee\gamma\gamma + \slashed{E}_T$ candidate event, as well as to search for GMSB SUSY. Both results optimize for GMSB scenarios. CDF searches in 202 pb$^{-1}$ of data for two photons with $E_T$>13 GeV, and $\slashed{E}_T$>45 GeV. As shown in Figure 11, a total of 0.27±0.12 events are expected with 0 events observed in the data (both updated since the conference). DØ searched in 185 pb$^{-1}$ of data in a slightly complementary fashion by requiring two photons with $E_T$>20 GeV and $\slashed{E}_T$>40 GeV. In this case, 2.5±0.5 events are expected with 1 event observed in the data. Since there is no excess in either case, both set limits on GMSB SUSY with limits from DØ shown in Figure 12. DØ sets limits on a lightest Chargino mass at 192 GeV, while CDF sets limits at 167 GeV. Unfortunately, there is still disagreement between the collaborations as to whether these two numbers may be directly compared. However, it is believed that the DØ limit, which has since been updated to 195 GeV [7], with a corresponding Neutralino limit of 108 GeV, is currently the new world's best.

We note that there is an interesting $e\gamma\gamma + \not{E}_T$ candidate event in the DØ data. See Figure 13. The photon $E_T$'s are 69 and 27 GeV respectively. There is an additional electron with 24 GeV of $E_T$ and all three objects are well measured and well separated. There is also $\not{E}_T = 63$ GeV in the event. Could this be $W\gamma\gamma$? A "cousin" of CDF $ee\gamma\gamma + \not{E}_T$ candidate? While a SUSY explanation is appealing, unfortunately most of the theoretically favored region to explain the $ee\gamma\gamma + \not{E}_T$ candidate is excluded. Clearly more data is needed.

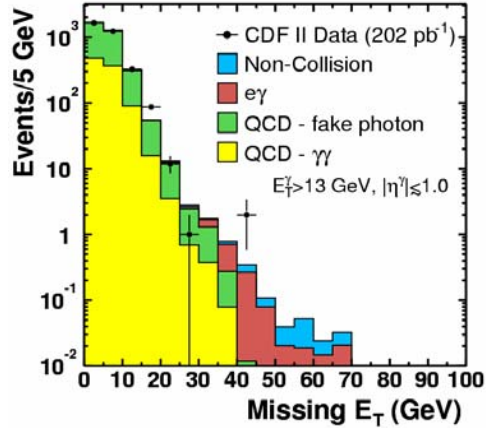

**FIGURE 11.** The $\not{E}_T$ distribution for the CDF $\gamma\gamma$ data sample (updated since the conference). There are no candidates above the $\not{E}_T > 45$ GeV threshold.

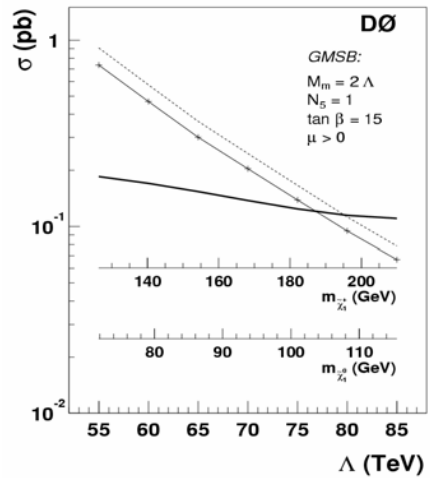

**FIGURE 12.** The 95% C.L. exclusion region from DØ on GMSB SUSY from the $\gamma\gamma + \not{E}_T$ channel. The exclusion of Charginos above 192 GeV is currently the world's best limit.

# CONCLUSIONS

It is clearly an exciting time to be at the Tevatron as it is the high-energy frontier for the next many years. Preliminary results on the first 200 pb$^{-1}$ are starting to come rapid-fire with first publications already submitted and accepted. Many of these results are the world's most sensitive results on the next 200 pb$^{-1}$ are in the pipeline and the detectors and the Tevatron continue to improve and provide new sensitivity. As shown in this talk, there are many interesting prospects that merit further attention as more data comes in.

## ACKNOWLEDGMENTS

The author would like to thank the many people for their assistance with the preparation of this talk. They include Richard Arnowitt, Alexander Belyaev, Volker Buescher, Ray Culbertson, Yuri Gershtein, Mike Gold, Jean-Francois Grivaz, Beate Heinemann, Teruki Kamon, Muge Karagoz-Unel, Min Suk Kim, Stephan Lammel, Sung-Won Lee, Jim Linnemann, Carsten Rott, Rick Snider, David Stuart, Song Ming Wang and Adam Yurkewicz.

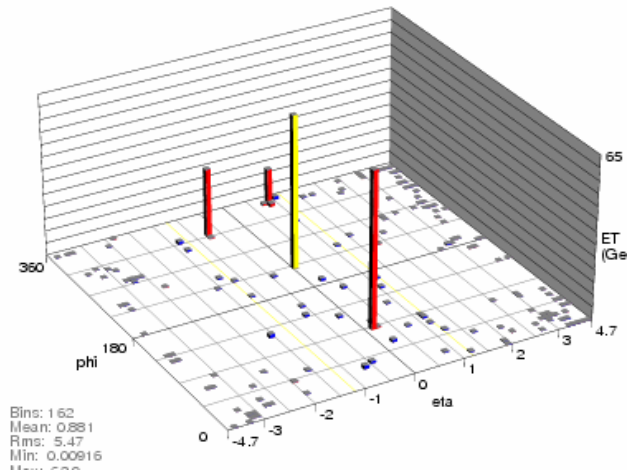

**FIGURE 13.** An interesting event from the DØ collaboration. This event has two photons, an electron and $\not{E}_T$. All are well measured and well separated. It is interesting to speculate if this event is related to the CDF ee$\gamma\gamma$+ $\not{E}_T$ candidate event.